\documentclass[prl,twocolumn,nofootinbib,showpacs,superscriptaddress,amsmath,amssymb,amsfonts]{revtex4}

\usepackage{bm}
\usepackage{color,xcolor}
\usepackage{mathtools,amsbsy}
\usepackage{keyval,graphicx}
\usepackage{slashed}
\usepackage{epstopdf}
\usepackage{isotope}
\usepackage{dcolumn}

\newcommand{\be}{\begin{equation}}
 \newcommand{\ee}{\end{equation}}
\newcommand{\ba}{\begin{array}{c}}
 \newcommand{\ea}{\end{array}}
\newcommand{\bea}{\begin{eqnarray}}
  \newcommand{\eea}{\end{eqnarray}}

\def\vec#1{\boldsymbol{#1}}

\begin{document}
\title{Double heavy tri-hadron bound state via delocalized {\boldmath$\pi$}  bond}

\author{Li Ma}\email{ma@hiskp.uni-bonn.de}
\affiliation{Helmholtz-Institut f\"ur Strahlen- und Kernphysik and Bethe
Center for Theoretical Physics, \\Universit\"at Bonn,  D-53115 Bonn, Germany}
\author{Qian Wang}\email{wangqian@hiskp.uni-bonn.de}
\affiliation{Helmholtz-Institut f\"ur Strahlen- und Kernphysik and Bethe
Center for Theoretical Physics, \\Universit\"at Bonn,  D-53115 Bonn, Germany}
\author{Ulf-G.~Mei{\ss}ner}\email{meissner@hiskp.uni-bonn.de}
\affiliation{Helmholtz-Institut f\"ur Strahlen- und Kernphysik and Bethe
Center for Theoretical Physics, \\Universit\"at Bonn,  D-53115 Bonn, Germany}
\affiliation{Institut f\"{u}r Kernphysik, Institute for Advanced
Simulation, and J\"ulich Center for Hadron Physics,\\
Forschungszentrum J\"ulich,  D-52425 J\"{u}lich, Germany}


\begin{abstract}
The number of  exotic candidates which are beyond the conventional quark model has grown dramatically 
during the last decades. Some of them could be
viewed as analogues of  the deuteron. Similarly, the existence of the
triton indicates that bound states formed by three hadrons could also exist.
 To illustrate this possibility, we study the $DD^*K$ and $BB^*\bar{K}$ systems by using the 
Born-Oppenheimer Approximation. To leading order, only one-pion exchange potentials are considered, 
which means that
the three constitutes share one virtual pion. That is similar to the role of the 
delocalized {\it $\pi$ bond} for the formation of Benzene in chemistry. 
 After solving the Schr\"odinger equation, we find two
three-body $DD^*K$ and $BB^*\bar{K}$ bound states with masses $4317.92_{-4.32}^{+3.66}~\mathrm{MeV}$ 
and $11013.65_{-8.84}^{+8.49}~\mathrm{MeV}$, respectively.  The masses of their 
$D\bar{D}^*K$ and $B\bar{B}^*\bar{K}$ analogues are $4317.92_{-6.55}^{+6.13}~\mathrm{MeV}$ and 
$11013.65_{-9.02}^{+8.68}~\mathrm{MeV}$. 
From the experimental side, the $D\bar{D}^*K$ bound state could be found by analyzing 
the current world data of the $B\to J/\psi\pi\pi K$ process by  focusing on the $J/\psi \pi K$ channel. 
Its confirmation could also help to understand 
the formation of  kaonic nuclei in nuclear physics.
 \end{abstract}
\date{\today}
\maketitle


The idea of hadronic molecules is largely motivated by the existence of the deuteron as a bound 
state of a proton and a neutron, for a recent review, see \cite{Guo:2017jvc}.
Thus, any development of nuclear physics could have an impact on hadron physics. An 
example would be the triton, a bound state of three nucleons, which raises the question  whether 
there are hadronic molecules formed by  three hadrons. Another indication from nuclear physics is 
the existence of possible kaonic nuclear bound state. 
The strong attraction of the $\bar{K}N$ systems leads to the bound state $\Lambda(1405)$~\cite{Dalitz:1959dn,Dalitz:1960du,Kaiser:1995eg,Oller:2000fj,Jido:2003cb,Ikeda:2012au,Guo:2012vv,Mai:2012dt,Cieply:2016jby}, especially to its two pole structure as reviewed by PDG~\cite{Olive:2016xmw}. This has led to speculations of 
 deeply bound kaonic states in light nuclei, i.e. $\bar{K}NN$~\cite{Akaishi:2002bg,Yamazaki:2002uh}.  
The confirmation of this kaonic light nuclei by E15 Collaboration in the $^3\mathrm{He}(K^-,\Lambda p) n$ process~\cite{Sada:2016nkb,Hashimoto:2014cri} adds further motivation 
for both nuclear and hadronic physicists to consider three-body bound states.  
On the other hand, it also indicates that there could be three-body bound states formed by three hadrons. 
A simple extension is replacing the nucleons by charmed mesons, i.e. the $D^{(*)}D^{(*)}K$ system.

The study of the $D^{(*)}D^{(*)}K$ system in both theoretically and experimentally
 could not only help to understand its two-body subsystems, but also to get insight into
  kaonic light nuclei.
\begin{figure}
\begin{center}
  \includegraphics[width=0.2\textwidth]{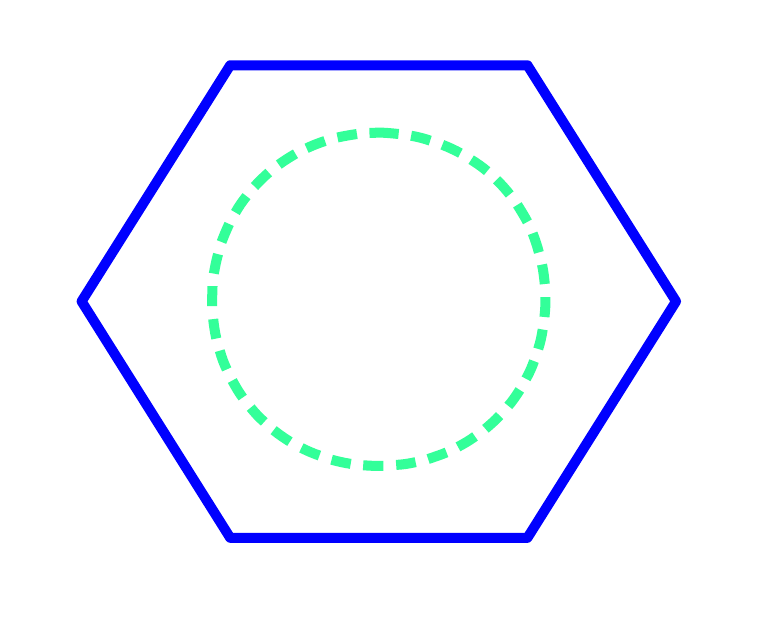} 
\caption{A simplified illustration of a benzene ring as a hexagon with a circle describing  the delocalized {\it $\pi$ bond} inside.}
 \label{fig:Benzene}
\end{center}
\vspace{-7mm}
\end{figure}
To the leading order, the behaviour that three constitutes share one virtual pion is similar to the idea of the delocalized
{\it $\pi$ bond} for the formation of Benzene in molecular physics as shown in Fig.~\ref{fig:Benzene}. 
Although the investigation of the three-body system in hadron physics have been an important topic of hadron physics for quite a while, 
 most of the calculations are in momentum space, such as the $\phi K\bar{K}$~\cite{MartinezTorres:2008gy}, $KK\bar{K}$~\cite{Torres:2011jt}, $f_0(980)\pi\pi$~\cite{MartinezTorres:2011vh}, $J/\psi K\bar{K}$~\cite{MartinezTorres:2009xb}, $DKK(\bar{K})$~\cite{Debastiani:2017vhv},
$BD\bar{D}$, $BDD$~\cite{Dias:2017miz} and $B^{(*)}B^{(*)}\bar{B}^{(*)}$~\cite{Wilbring:2017fwy} mesonic systems and the $DNN$~\cite{Bayar:2012dd}, $NDK$, $\bar{K}DN$, $ND\bar{D}$~\cite{Xiao:2011rc}, and $N\bar{K}K$~\cite{Jido:2008kp} baryonic systems.
Alternatively, we deal with the three-body problem in coordinate space by solving the Sch\"odinger equation, 
because one can extract the size of the bound state directly,
 which could be used as a  criterion  whether hadrons are the effective degrees of freedom or not.

\begin{figure*}
\begin{center}
  \includegraphics[width=0.99\textwidth]{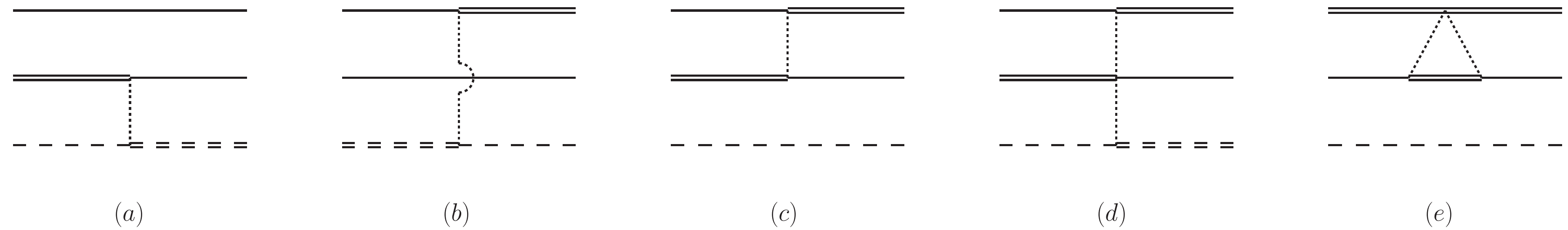}
\caption{Diagrams (a), (b) and (c) are the leading One-Pion-Exchange (OPE) diagrams
 for the transitions among the relevant three-body channels, 
i.e. $DD^*K$, $DDK^*$ and $D^*DK$ channels.
These three channels are labeled as the first, second and third channels, respectively. Thus, the diagrams (a), (b), (c) indicate the transition potentials $V_{12}$, $V_{23}$ and $V_{13}$, in order. The (double) solid, dashed lines represent the $D^{(*)}$, $K^{(*)}$ fields.
Dotted lines denote pion fields. The Two-Pion-Exchange (TPE) diagrams, i.e. (d) and (e), are the next-to-leading order contributions.} \label{fig:FeyD}
\end{center}
\vspace{-5mm}
\end{figure*}
In this letter, we solve the three-body Schr\"odinger equation
to study whether there is a bound state for the $DD^*K$ system. 
As the kaon mass is considerably  smaller than those of the charmed mesons, 
the Born-Oppenheimer (BO) approximation can be applied to simplify this case, 
 but with an uncertainty of the order $\mathcal{O}(m_K/m_{D^{(*)}})$.
 The procedure
 is divided into two steps. Firstly, we keep the two heavy mesons, i.e. $D$ and $D^*$,
 at a given fixed location $R$ and study the dynamical behaviour of the light kaon. 
 In the next step, we solve Schr\"odinger equation of the $DD^*$
 system with the effective BO potential created from the interaction with the kaon. 
 
 As the Two-Pion-Exchange (TPE) diagrams (Figs.~\ref{fig:FeyD} (d) and (e)) are
  the next-to-leading order contribution similar to that in nuclear physics~, see e.g. 
 Refs.~\cite{Epelbaum:2008ga,KalantarNayestanaki:2011wz,Hammer:2012id}  for reviews, 
 we only consider the leading order One-Pion-Exchange (OPE) diagrams, 
 i.e. Figs.~\ref{fig:FeyD} (a), (b) and (c), which is analogous to the delocalized {\it $\pi$ bond} in molecular physics.
 That is because there is only one virtual pion shared by the three constituents, as shown in Figs.~\ref{fig:FeyD} (a), (b) and (c), instead of localizing between any two of them. Note that it is important
to consider these diagrams together. In this sense,
  this behaviour is similar to that three pairs of electrons are shared by the six carbon atoms in Benzene.
 As a result, we work within the framework which respects $\mathrm{SU}(2)$ flavor symmetry
 \footnote{The Lagrangian in Ref.~\cite{Cleven:2010aw} is under the $\mathrm{SU}(3)$ symmetry where the interactions of $DK$ and $D^*K$ are the same to the leading order.}.  
 The relevant Lagrangian is  
 \begin{eqnarray*}
&\mathcal{L}&=-i\frac{2g_P}{F_\pi}\bar{M}
P^{*\mu}_b\partial_{\mu}\phi_{ba}P^{\dag}_{a}+i\frac{2g_P}{F_\pi}\bar{M} P_b\partial_{\mu}\phi_{ba}P^{*\mu\dag}_{a} 
\label{pseudo-exchange}
\end{eqnarray*}
with $P^{(*)}=(D^{(*)0}, D^{(*)+})$ or $(K^{(*)-},\bar{K}^{(*)0})$ and
\begin{eqnarray}
\phi=\left(
         \begin{array}{cc}
           \frac{\pi^0}{\sqrt{2}} & \pi^+ \\
           \pi^- & -\frac{\pi^0}{\sqrt{2}}\\
         \end{array}
       \right).
\end{eqnarray}
Here, $\bar{M}=\sqrt{M_{P}M_{P^*}}$ is the difference of the normalization factor between the relativistic and non-relativistic fields.
 The effective couplings $g_{D}$ ($g_K$)
can be extracted from the partial width of the $D^*\to D\pi$ ($K^*\to K\pi$) process. 
As there is no direct OPE diagram for the $D^*K$ channel, an additional channel $DK^*$
\footnote{The reason why $D^*K^*$ channel is not included is because it is the next higher threshold.} is included
as an intermediate channel.  Since the three momentum $p_K$ of the $D^*K$ system at the $DK^*$ threshold is 
$61\%$ of its reduced mass $\mu_{D^*K}$, the relativistic effect of the kaon is not negligible. That is the reason why
we use the relativistic form of the interaction and keep the expression for the kaon to 
the order  $\mathcal{O}(\frac{p_K}{m_K})$.
 To take into account the substructure for each pion vertex, a monopole form factor $\mathcal{F}(q)=\frac{\Lambda^2-m_\pi^2}{\Lambda^2-q^2}$, with $q$ the four-momentum of the pion and $\Lambda$ the cutoff parameter. With the cutoff $\Lambda_{D^*K}=803.2~\mathrm{MeV}$,
we find a $D^*K$ bound state with mass at $D_{s1}(2460)$. The dynamics of the $I=\frac 12$ $DD^*K$ system
with an isosinglet $D^*K$ does not depend on the cutoff parameter for the $DD^*$ system whose attractive and repulsive parts are cancelled exactly. 
\begin{figure*}
\begin{center}
  \includegraphics[width=0.34\textwidth]{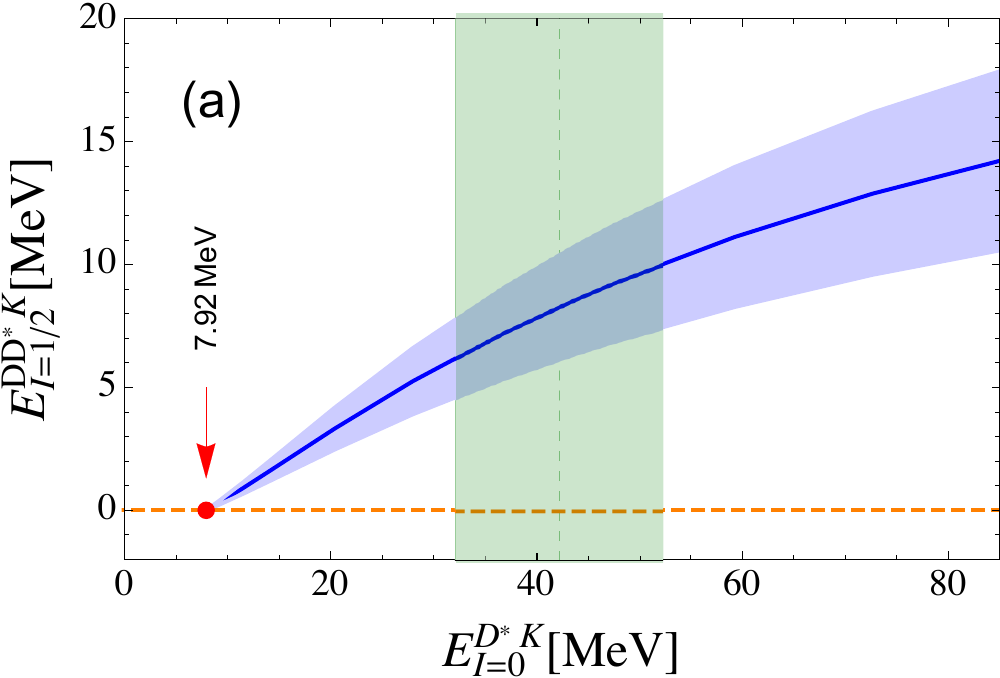}  \quad \includegraphics[width=0.35\textwidth]{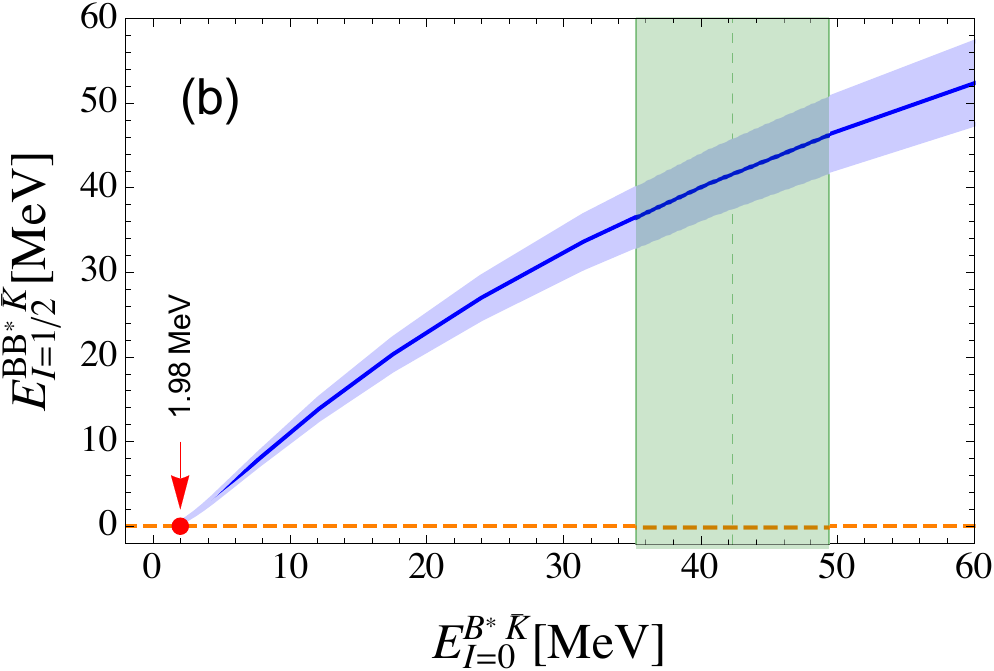}  
\caption{The binding energy (defined in Eq.~\eqref{eq:BE3}) of the $DD^*K$ three-body system with $I=1/2$ in terms of that of the isosinglet $D^*K$ two-body system is presented on the left panel. The uncertainty is estimated as $m_K/(2\mu_{DD^*})$. The right panel shows the corresponding dependence for the $BB^*\bar{K}$ system. The red point is the critical point which indicates the lower limit of the required binding energy of the isosinglet $D^*K$ or $B^*\bar{K}$ to form a three-body bound state. 
The vertical dashed lines and bands are the central values and uncertainties of the binding energies of the two-body subsystems from the chiral dynamical analysis~\cite{Guo:2006rp}.} \label{fig:BindingEnergy}
\end{center}
\vspace{-7mm}
\end{figure*}
\begin{figure}
\begin{center}
   \includegraphics[width=0.25\textwidth]{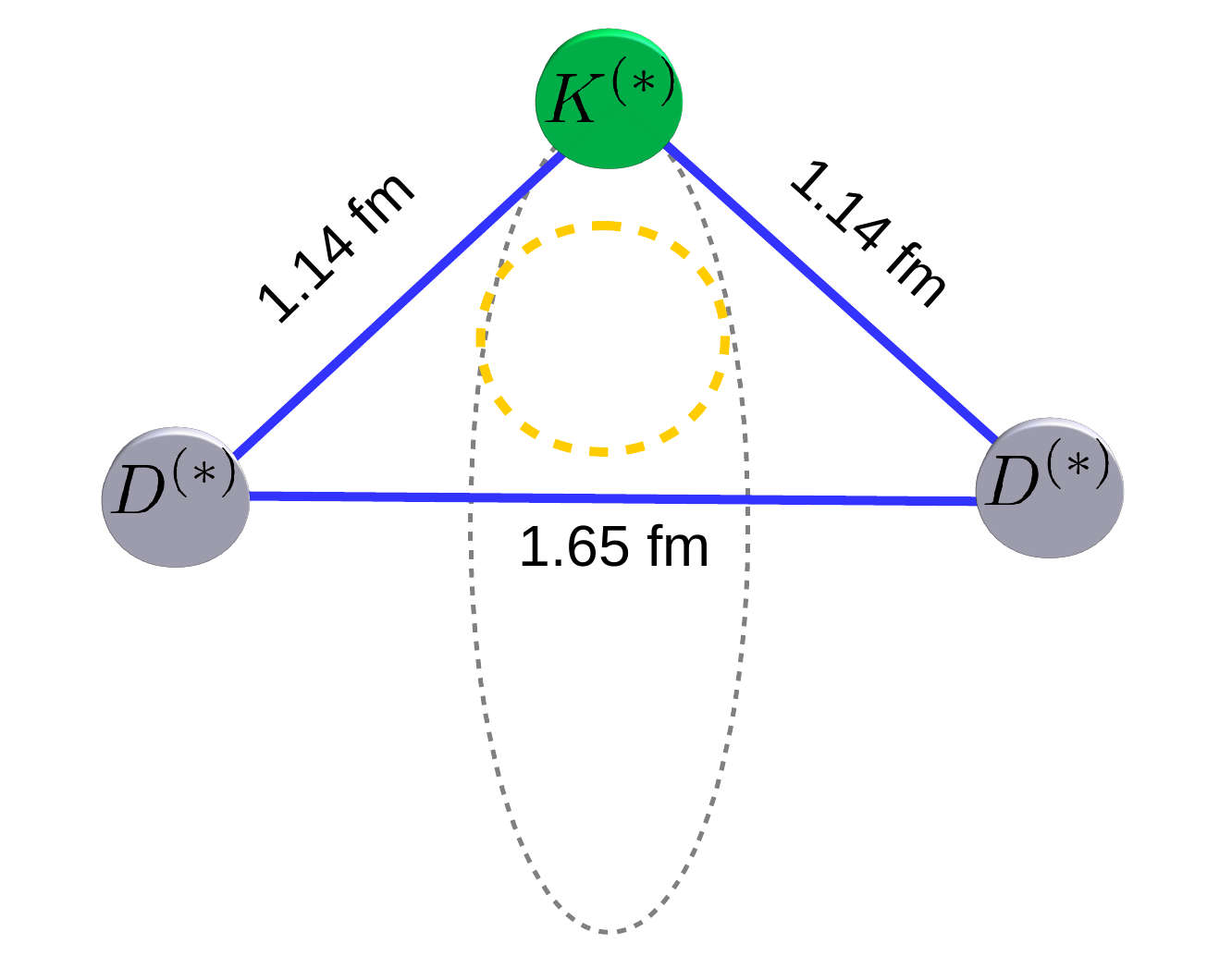}
\caption{Formation of the three-body $DD^*K$ bound state through the delocalized {\it $\pi$ bond}. 
The root-mean-square radius of each two-body subsystem is indicated explicitly.} \label{fig:RadiusD}
\end{center}
\vspace{-5mm}
\end{figure}

The BO approximation is based on the factorized wave function 
\begin{eqnarray}
|\Psi_T(\vec{R},\vec{r})\rangle=|\Phi(\vec{R})\Psi(\vec{r_1},\vec{r_2})\rangle~,
\end{eqnarray}
with the two charmed mesons and the light kaon located at $\pm \vec{R}/2$ and $\vec{r}$. Here, $\vec{r}_1=\vec{r}+\vec{R}/2$
and $\vec{r}_2=\vec{r}-\vec{R}/2$ are the coordinates of the kaon relative to the first and second interacting $D^*$.  In our case,
due to the OPE potential, the three channels, i.e. $DD^*K$, $DDK^*$ and $D^*DK$, are coupled with each other as shown
 in Figs.~\ref{fig:FeyD} (a), (b), (c). 
The wave function of kaon in the $DD^*K$ system is the superposition of the two two-body subsystems
\begin{eqnarray*}
|\Psi(\vec{r}_1,\vec{r}_2)\rangle &=& C_0\{ \psi (\vec{r}_2)~|D D^{\ast} K\rangle + \psi (\vec{r}_1)~|D^{\ast} D K\rangle\\\nonumber
&+&C[\psi' (\vec{r}_1)+\psi' (\vec{r}_2)]|D D K^{\ast}\rangle\}~.
\end{eqnarray*}
The constant $C_0$ can be fixed by the normalization constraint of the total wave function.

\begin{figure}
\begin{center}
 \includegraphics[width=0.18\textwidth]{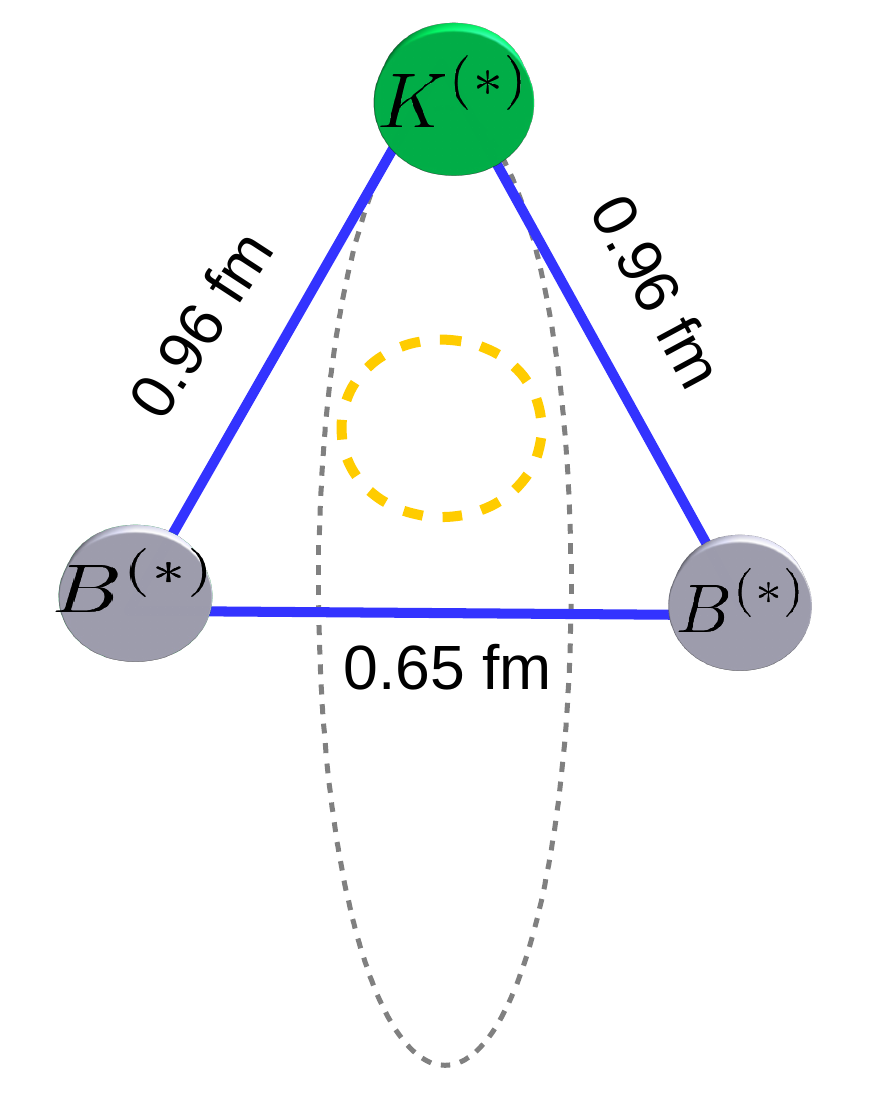}
\caption{The same as that in Fig.~\ref{fig:RadiusD}, but for the $BB^*\bar{K}$ system.} \label{fig:RadiusB}
\end{center}
\vspace{-5mm}
\end{figure}

In the BO approximation, the three-body Sch\"odinger equation can be simplified into two sub-Sch\"odinger equations. 
One is the equation for the kaon
\begin{eqnarray}
H(\vec{r}_1,\vec{r}_2)|\Psi(\vec{r}_1,\vec{r}_2)\rangle=E_K(R)|\Psi(\vec{r}_1,\vec{r}_2)\rangle
\label{eq:BE3}
\end{eqnarray}
at any given $\vec{R}$ with 
\begin{eqnarray}\nonumber
H(\vec{r}_1,\vec{r}_2)=\left(
         \begin{array}{ccc}
           T_{11}(\vec{r}_1,\vec{r}_2) & V_{12}(\vec{r}_2) & 0  \\
           V_{21}(\vec{r}_2) & \delta M+T_{22}(\vec{r}_1,\vec{r}_2)  & V_{23}(\vec{r}_1) \\
          0 & V_{32}(\vec{r}_1) & T_{33}(\vec{r}_1,\vec{r}_2) \\
         \end{array}
       \right).
\label{eq:ham}
\end{eqnarray}
Here, $T_{ii}$ is the relative kinetic energy for the $K$ in the $i$th channel and $\delta M=M_{D}+M_{K*}-M_{D*}-M_{K}$ is the mass gap between the $DDK^*$ and $DD^*K$ ($D^*DK$) channels. 
The parameter $C$ can extracted from the variation principle ${\partial E_{K}(R)}/{\partial C}=0$.
The other one is the Sch\"odinger equation for the two heavy charmed mesons
\begin{eqnarray}
H^\prime(\vec{R})|\Phi(\vec{R})\rangle=-E_3|\Phi(\vec{R})\rangle
\end{eqnarray}
with 
\begin{eqnarray}\nonumber
H^\prime(\vec{R})&=&T_h(\vec{R})+V_h(\vec{R})+V_{\rm BO}(\vec{R})\\\nonumber
&=&\left(
         \begin{array}{ccc}
         T_{DD^{\ast}}(\vec{R}) & 0  & V_{13}(\vec{R})  \\
 0 & T_{DD}(\vec{R})  & 0  \\
 V_{31}(\vec{R}) & 0  & T_{DD^{\ast}}(\vec{R)} 
         \end{array}
       \right)\\\nonumber
&+&V_{\rm BO}(\vec{R})
\end{eqnarray}
where $V_{BO}(\vec{R})=E_K(\vec{R})+E_B$ is the BO potential provided by the kaon and $E_B$ is 
the binding energy of the isosinglet $D^*K$ system. The total energy of the three-body system relative to 
the $DD^*K$ threshold is $E=-(E_3+E_B)$. 
Since the three-body force only appears at next-to-leading order, which is neglected here,
the dynamics of the three-body system can be described by that of its  two-body sub-systems. 
Since the potential of the isospin singlet (triplet) 
$D^*K$ is attractive (repulsive), only the three-body $DD^*K$ system with total isospin $\frac 12$ and isosinglet $D^*K$ could form a bound state 
\begin{eqnarray}\nonumber
|DD^*K\rangle_{\frac{1}{2},\frac{1}{2}} &=& \frac{1}{\sqrt{2}}[| D^{+}(D^{\ast +}K^{0})_0 \rangle+| D^{+}(D^{\ast 0}K^{+})_0 \rangle],  \\\nonumber
|DD^*K\rangle_{\frac{1}{2},-\frac{1}{2}} &=& \frac{1}{\sqrt{2}}[-| D^{0}(D^{\ast +}K^{0})_0 \rangle-| D^{0}(D^{\ast 0}K^{+})_0 \rangle],
\end{eqnarray}
where the subscripts denote the isospin and its third component. 
How deep it is depends on how strong the $D^{*}K$ attraction is. Fig.~\ref{fig:BindingEnergy}(a) shows the dependence of 
the energy $E_3$ of the $I(J^P)=\frac 12(1^-)$ $DD^*K$ system on the binding energy $E_B$ of its isosinglet
 two-body $D^*K$ system. The analogous bottom system is shown 
in Fig.~\ref{fig:BindingEnergy}(b). The two vertical bands in Fig.~\ref{fig:BindingEnergy} are the binding energies of the $D^*K$ and $B^*\bar{K}$ in 
Ref.~\cite{Guo:2006rp}, which allows one to deduce the energies $E_3$ of the two systems 
\begin{eqnarray}\nonumber
E_{I=1/2}^{DD^*K}=8.29^{+4.32}_{-3.66}~\mathrm{MeV},\quad E_{I=1/2}^{BB^*\bar{K}}=41.76^{+8.84}_{-8.49}~\mathrm{MeV},
\end{eqnarray}
where the uncertainties are estimated as $m_K/(2 \mu_{DD^*})$ and $m_K/(2 \mu_{BB^*})$ as discussed above. 
That means, after including the binding energies of the $D^*K$ and $B^*\bar{K}$ subsystems,
 there are two bound states with masses $4317.92_{-4.32}^{+3.66}~\mathrm{MeV}$ and $11013.65_{-8.84}^{+8.49}~\mathrm{MeV}$,
  respectively. The critical point in Fig.~\ref{fig:BindingEnergy} 
refers to the case that when the binding energy of the $D^*K$ or $B^*\bar{K}$ is larger than that value, a three-body bound state starts to emerge. 

\begin{figure}
\begin{center}
  \includegraphics[width=0.4\textwidth]{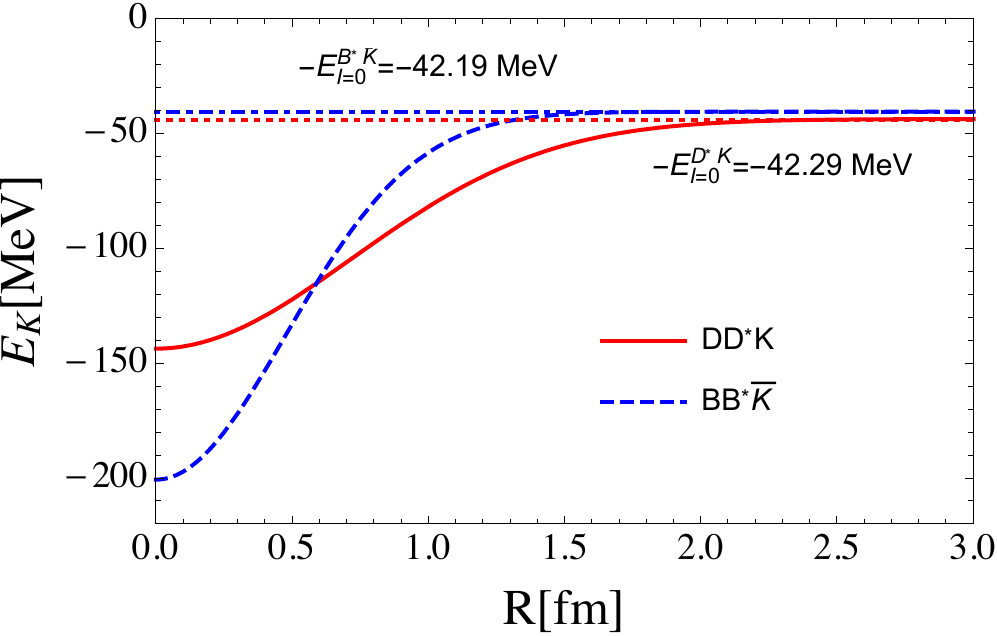}  
\caption{The kaon energy in the three-body system 
as a function of the distance $R$ between two heavy-light mesons. The red dotted and blue dot-dashed horizontal lines are 
the binding energies of the isosinglet $D^*K$ and $B^*\bar{K}$ systems, respectively. When the distance $R$ is larger than a certain value,
 the kaon energy of the three-body system equal to the binding energy of the isosinglet $D^*K$ or $B^*\bar{K}$ two-body system. 
 The two-body binding energies are from Ref.~\cite{Guo:2006rp}.} \label{fig:totalenergy}
\end{center}
\end{figure}

The corresponding root-mean-square radii for each two-body subsystem are shown in 
Figs.~\ref{fig:RadiusD} and \ref{fig:RadiusB} for the charm and bottom sectors, respectively. 
The root-mean-square between $K$ and $D^{(*)}$ is $1.14~\mathrm{fm}$. That value for 
the $DD^*$ is $1.65~\mathrm{fm}$. 
Since the $S$-wave component dominates the isosinglet $D^*K$ wave function, the kaon is  
distributed evenly on a circle with a  root-mean-square radius of $0.79~\mathrm{fm}$  
on the plane perpendicular to the $DD^*$ direction.  
The case for the bottom sector is similar
as shown in Fig.~\ref{fig:RadiusB}. The root-mean-square radius between the kaon and bottom mesons is $0.96~\mathrm{fm}$. 
The distance between a pair of bottom mesons is $0.65~\mathrm{fm}$, which indicates that the kaon is distributed on a circle with a root-mean-square radius of $0.90~\mathrm{fm}$
on the plane perpendicular to the $BB^*$ direction. All the above root-mean-square radii are large 
enough to separate the relevant 
two constitutes, making hadrons the effective degrees of freedom. Especially, 
although the root-mean-square radius between $BB^*$ is as small as $0.65~\mathrm{fm}$,
it is still much larger than two times of the Compton wave lengths of  the bottom meson. 

The kaon energy in the three-body system defined by Eq.~\eqref{eq:BE3} 
 in terms of the distance between the two heavy mesons is shown in Fig.~\ref{fig:totalenergy}. 
When the distance goes to infinity, it goes back to the binding energy of the two-body subsystem. 
Furthermore, when the charmed (bottomed) system gains another $42.29~\mathrm{MeV}$ ($42.19~\mathrm{MeV}$) energy,
the three-body system will totally  break up into three individual particles. The dependence of bottomed system (blue dashed curve)
is narrower, but deeper, which means that its size is smaller, but its binding energy is larger. That is consistent with what we obtained above.

As the long-distance $DD^*$ potential from the OPE is related to that of the $D\bar{D}^*$ potential
 by $G$-parity~\cite{Klempt:2002ap}, there could also exist a three-body $D\bar{D}^*K$ bound state, 
 but with additional uncertainty $m_\pi^2/(2\mu_{DD^*})$ (which characterizes the natural energy scale of OPE~\cite{Braaten:2007dw})
  from the unknown short-distance interaction.  
 Thus, for the $D\bar{D}^*K$ and $B\bar{B}^*\bar{K}$ system, the three-body binding energies are
\begin{eqnarray}\nonumber
E_{I=1/2}^{D\bar{D}^*K}&=&8.29^{+6.55}_{-6.13}~\mathrm{MeV}, ~E_{I=1/2}^{B\bar{B}^*\bar{K}}=41.76^{+9.02}_{-8.68}~\mathrm{MeV}
\end{eqnarray}
with the additional uncertainty from the missing short-distance interaction. Those correspond to two bound states with masses 
$4317.92_{-6.55}^{+6.13}~\mathrm{MeV}$ and $11013.65_{-9.02}^{+8.68}~\mathrm{MeV}$. 
For the $I=1/2$ $D\bar{D}^*K$ three-body bound state
\begin{eqnarray}\nonumber
|D\bar{D}^*K\rangle_{\frac 12, \frac 12} &=& \frac{1}{\sqrt{2}}[-| D^{+}(\bar{D}^{\ast 0}K^{0})_0 \rangle+| D^{+}(D^{\ast -}K^{+})_0 \rangle],\\\nonumber
|D\bar{D}^*K\rangle_{\frac 12, -\frac 12} &=& \frac{1}{\sqrt{2}}[| D^{0}(\bar{D}^{\ast 0}K^{0})_0 \rangle-| D^{0}(D^{\ast -}K^{+})_0 \rangle],  
\end{eqnarray}
 as the total isospin of $D^*K$ is $0$, the fraction of
  the isospin triplet $D\bar{D}^*$ in the bound state is three times as large as that of the isospin singlet one. 
Thus, the most easy channel to detect it is the $J/\psi \pi K$ channel. One could also notice
that the three-body $D\bar{D}^*K$ ($B\bar{B}^*\bar{K}$) bound states have either neutral or positive (negative) charge.
Aiming at the $X(3872)$, LHCb, Belle and BABAR have collected quite numerous data for $B$ decays in the $J/\psi\pi\pi K$ channel. However, they focus on the $J/\psi\pi\pi$ channel.  
The existence of the $D\bar{D}^*K$ bound state could be checked
 from the experimental side by analyzing the current world data on the channels
$J/\psi\pi^+ K^0$, $J/\psi\pi^0 K^+$, $J/\psi\pi^0 K^0$, and $J/\psi\pi^- K^+$. 
As the charged particle is most easily  detectable in experiment, the last channel is the 
most promising  one to search for the new state. 

As a short summary, the dynamics of the three-body system can be reflected by that of its two-body subsystem to leading order.
Based on the attractive force of the isosinglet $D^*K$ and $B^*\bar{K}$ systems, 
we predict that there exist two $DD^*K$ and $BB^*\bar{K}$ bound states
 with $I=\frac 12$ and masses $4317.92_{-4.32}^{+3.66}~\mathrm{MeV}$ and $11013.65_{-8.84}^{+8.49}~\mathrm{MeV}$, respectively.
  The $D\bar{D}^*K$ and $B\bar{B}^*\bar{K}$ systems are their analogues with masses 
  $4317.92_{-6.55}^{+6.13}~\mathrm{MeV}$ and $11013.65_{-9.02}^{+8.68}~\mathrm{MeV}$, where the additional uncertainties  
stemming from the unknown short-distance interaction.  The existence of the $D\bar{D}^*K$ bound state
could be checked from the experimental side by  analyzing the data
for the $B\to J/\psi\pi\pi K$ channel, by focusing on the $J/\psi \pi K$ channel. Its 
confirmation could also help to understand 
the formation of the kaonic nuclei in nuclear physics.


\medskip

We are grateful to Johann Haidenbauer, Jin-Yi Pang and Akaki G. Rusetsky,
 for the useful discussions, and especially Meng-Lin Du and Jia-Jun Wu.
We acknowlegde contributions from Martin~Cleven during the early stage of this investigation.
This work is
supported in part by the DFG (Grant No. TR110) and the NSFC (Grant No. 11621131001) through funds 
provided to
the Sino-German CRC 110 ``Symmetries and the Emergence of Structure
in QCD''. The work of UGM was also supported by the Chinese Academy 
of Sciences (CAS) President's International Fellowship Initiative (PIFI) 
(Grant No. 2017VMA0025).

\end{document}